\documentclass{article}
\usepackage[utf8]{inputenc}

\title{Comments Regarding `On the Identifiability of the Influence Model for Stochastic Spatiotemporal Spread Processes'}
\author{Sandip Roy\thanks{Correspondence should be sent to sandip@wsu.edu .  This work was partially supported by United States National Science Foundation grant CNS-1545104.}}
\date{November 4, 2018}

\usepackage{graphics}
\usepackage{amsmath}
\usepackage{amssymb}
\usepackage{epsfig,wrapfig}
\usepackage{graphicx}
\usepackage{amsmath}
\usepackage{epsfig,wrapfig}
\usepackage{color}
\usepackage{cite}
\usepackage{fullpage}

\begin{document}

\maketitle

\begin{abstract}
The identifiability analysis of a networked Markov chain model known as the influence model, as described in \cite{he}, is examined.  Two errors in the identifiability analysis -- one related to the unidentifiability of the partially-observed influence model, the second related to an omission of an additional recurrence criterion for identifiability -- are noted.  In addition, some concerns about the formulation of the identifiability problem and the proposed estimation approach are noted.  
\end{abstract}

\section{Introduction}

The {\em influence model}, which describes a class of networked discrete-state Markov chains with quasi-linear interactions \cite{asa1,asa2}, has proved useful for representing social processes in human groups \cite{pentland1}, environmental phenomena (e.g. convective weather propagation) \cite{xue1,xue2}, and decision-making algorithms \cite{wan}, among other stochastic network dynamics.  The model is appealing for myriad applications because of: 1) the tractability enabled by its quasi-linear structure, and 2) the model's ability to represent high-dimensional stochastic network processes with a terse parameterization.  In particular, the terseness of the model description suggests that model learning may be possible using limited data.  With this motivation, several studies (e.g. \cite{pentland1,xue1,xue2}) have proposed methods for estimating influence model parameters from observed data. However, these various studies have provided procedures and heuristics for parameter estimation, rather than formal guarantees about estimability of the parameters.  

Very recently, an ambitious study by C. He and co-workers has sought to establish conditions for the {\em identifiability} of the influence model \cite{he}, i.e. conditions under which almost-sure estimation of model parameters is achieved in the large-data limit.  The study characterizes identifiability for both the case that the influence model is fully observed, and the case where only a subset of the sites (chains) in the network are observed. Additionally, the study proposes some new algorithms for the estimation of influence model parameters. \newline

This comment firstly notes two technical flaws in \cite{he}:

1) The identifiability conditions for the partially-observed influence model (POIM) are based on Markov-chain model for the observed sites' joint status (see pages 5 and 14-16 of \cite{he}).  The authors claim  that the state transition matrix for the observed sites' joint status can be computed as a projection of the {\em master Markov chain} of the influence model, which tracks the joint status of all sites in the network. They then assert that that the POIM is necessarily unidentifiable since the master Markov chain cannot be uniquely determined even if the reduced Markov model is exactly computed. However, the observed sites' joint status is not generally governed by a Markov process.  For instance, consider a partially observed influence model with network matrix $D=\begin{bmatrix} 0.6 & 0.4 \\ 0.3 & 0.7 \end{bmatrix}$.  and local transition matrix $A=\begin{bmatrix} 0.9 & 0.1 \\ 0.2 & 0.8 \end{bmatrix}$, for which only the first site is measured.  The master Markov chain's transition matrix for this POIM, per
Equation (6) in \cite{he}, is $G=\begin{bmatrix} .81 & .09 & .09 & .01 \\ .2542 & .3658 & .1558 & .2242 \\ .3312 & .1488 & .3588 & .1612 \\ .04 & .16 & .16 & .64 \end{bmatrix}$.  For this model, one can check that the conditional probability 
$P(s_1[2]=1\, | \, s_1[1]=1)=0.8355$, provided that the model's initial state is governed by the master Markov chain's stationary distribution.  In contrast, $P(s_1[2]=1 \, | \, s_1[1]=1,s_1[0]=0)=0.7687$.  Thus, it is seen that the observed statuses of the POIM do not evolve in a Markov fashion.  Hence, the computation of a  reduced-dimension transition matrix in Equations (37) and (38) of \cite{he} is moot, and indeed no such reduced transition matrix can be defined  that captures the full stochastic evolution of the observed sites. In consequence, the core of the argument for unidentifiability (Theorem 12) -- namely, the master Markov chain's transition matrix cannot be recovered from the reduced transition matrix -- is flawed.  Indeed, it is easy to check that the master Markov chain for the example POIM can be estimated from the observation sequence, as the POIM is a hidden Markov model which satisfies the standard regularity conditions needed for identification \cite{petrie}; for this example, it then follows that the influence model's parameters can be identified.  The hidden Markov model perspective clarifies that the hidden sites' statuses incur an eventual impact on the measured sites (in a statistical sense), in a way that potentially can enable identification. 

2) The entries in the transition matrix of a Markov chain can be estimated almost surely using state trajectory data only if the Markov chain is recurrent, i.e. comprises a single recurrent class.  Lemma 1 and all ensuing identifiability results for the full state observation case (Corollary 1, Theorem 8, Corollary 3, Theorem 10), need to be revised to include this additional criterion.  In particular, we note that Corollary 3 -- which states that the binary influence model is always identifiable -- is invalidated by the additional requirement of recurrence.  The master Markov chain for the binary influence model is not recurrent (it necessarily has two absorbing states, corresponding to the two network-wide consensus states).  For this reason, the binary influence model is in fact generically not identifiable. For other classes of influence models, the master Markov chain may or may not be recurrent, see \cite{asa1}. \newline

Secondly, this comment identifies two concerns regarding the formulation of the identifiability problem in \cite{he}, and the techniques introduced for estimation of influence-model parameters:

1) The identifiability analysis for the fully-measured influence model is of questionable value.  When all sites in the influence model are observed, identifiability is lost only if influence models with different parameters are statistically identical.  Differentiation among statistically identical models is of little value, however, as simulation of any such model will produce statistically identical state trajectories.  The identifiability question is of interest if parts of the dynamics are hidden, and/or observations are subject to noise.

2) Some of the estimation techniques proposed by the authors are based on first estimating the master Markov chain's transition matrix, and computing the influence model's parameters thereof.  While this approach is helpful for proving identifiability, the resulting algorithms may not be practical.  The master Markov chain is very high dimensional, and estimation of its parameters may require a large data set.  Given that the influence model has a terse parameterization, techniques for directly estimating these parameters are appealing from a data-theoretic and computational standpoint. \newline

It is important to stress that, while the article \cite{he} has some technical deficiencies as noted above, it introduces a promising and challenging direction of research on identification of quasi-linear stochastic network models.  The article also provides a thorough review of the literature on influence models, and scopes the identification problem relative to the literature.
The authors are applauded for the exciting new perspective brought forth in their study.

\end{document}